# Compton Effect in the Medium with Non-unity Refractive Index[1]

S. G. Chefranov[1,*], A. G. Chefranov[2,**], and Vinay Venugopal[3,***]

[1]*A. M. Obukhov Institute of Atmospheric Physics, RAS, Moscow, Russia*

[2]*Eastern Mediterranean University, Famagusta, North Cyprus*

[3]*Division of Physics, VIT University, Chennai Campus, Chennai, India*

*Corresponding Author. Email: schefranov@mail.ru

**Email: Alexander.chefranov@emu.edu.tr

***Email: vinayvenugopal@vit.ac.in

26 July 2013

**Abstract**

The foundations of the modern quantum theory of electro-magnetic radiation and its interaction with medium are constituted by the Einstein's theory of photo-effect and the theory of Compton effect. However, in the theory proposed by Compton and subsequent refinements, the medium features in explicit form are not generally accounted at all. This theory is based on the relationship between the photon momentum and its energy that may take place only in vacuum, but not in a medium with non-unity refractive index $n$. Here, we use Abraham's representation for the momentum of a photon in medium for the creation of a new consistent Compton effect theory accounting for media features. A new condition under which anomalous Compton effect is possible, is defined. In our case, due to the photon scattering on a moving electron, the photon's energy can increase when the electron velocity is above a threshold value, in analogy with the Vavilov-Cherenkov radiation effect. In our theory we also provide a new description of the parameters for scattered electrons, which is important for the experimental determination of the dependence of the Compton effect in medium on refractive index $n$.

## 1. INRODUCTION

Currently, the widely used electromagnetic field (EMF) theory in the medium is based on Minkowski's theory [1] and its conclusions although more than forty years ago it's alternative,

---

[1] Accepted to PIERS 2013 in Stockholm, Sweden, 12-15 August, 2013

Abraham's theory [2], was confirmed experimentally. Actually, the experiments [3-5] could have ended the discussion that is on for the past sixty years on the so called Abraham-Minkowski dilemma since "the question is definitely solved on the benefit of Abraham's tensor" (see [6]), and Abraham's theory was accepted as the true one [6]. However, surprisingly Minkowski's theory up to now is not accepted as the false one and it continues to be used as being "more convenient and even slightly differing from Abraham's theory in the most cases" [6,7]. In the base of such positive relation to Minkowski's theory is, as mentioned in [6,7], its success in application of its conclusions to the building of quantum Vavilov-Cherenkov radiation (VCR) theory suggested by V.L. Ginzburg [6] in 1940 (see respective references in [6]), i.e. long before the experimental confirmation of Abraham's theory. Moreover, prior to appearance of the works [8-10], even attempts were not made based on Abraham's theory to create a quantum theory of VCR and its analogues including Compton effect in the medium theory. To do it, in [8-10], it was necessary to revise significantly the role of the medium energy change when a quantum of VCR is emitted, in spite of the fact that medium consideration as a direct VCR radiator was well known [11,12]. Such accounting of the medium energy change while medium emitting a VCR quantum is not conducted in [6] and, more over, it is impossible because of the complex value of effective photon rest mass which corresponds to using in [6] a photon momentum in the medium representation based on the Minkowski's theory. When using in [8-10] (see also [13-15]) a photon momentum in the medium representation in a form corresponding to the Abraham's theory, there is no such difficulty since in this case photon in the medium effective rest mass value is real and has physical meaning. This, as it is shown in [8-10], is found to be important for the description of a non-stationary and non-equilibrium process defining physical mechanism and itself the threshold for VCR effect realization by the medium, contrary to the Tamm-Frank [11] VCR theory and corresponding to quantum VCR theory of Ginzburg which give only description of already stabilized equilibrium VCR field in the medium. The above mentioned "success" of Ginzburg quantum VCR theory is related in [6] with the presence of a correspondence of that VCR theory with initially stationary Tamm-Frank VCR theory [11]. This theories correspondence [6] and [10] now finds explanation on the base of conclusions of [16], where actually limits of the Minkowski theory applicability are established limiting the opportunity of its use only for the description of stabilized stationary electromagnetic processes with the medium participation.

Hence it is necessary to revise on the base of Abraham theory, all the theories for which the very non-stationary non-equilibrium effects of interaction of EMF and the medium may be essential. In the present work, accounting conclusions of [17] we propose a new Compton effect theory in the medium having refractive index n (n>1 or n<1) and the significant impact of that parameter on the experimentally measured values characterizing the effect is defined. Note that in [18], it is also proposed to account for the refractive index but without use of representation of photon momentum in the medium in the Abraham's form and without corresponding use of the finite and real photon rest mass in the medium with non-unity refractive index.

## 2. RESULTS AND DISCUSSION

Let us consider the energy-momentum balance equation for the Compton effect in the medium description. Energy-momentum balance equations are as follows:

$$\Delta E + E_1 + E_i = E_2 + E_r \quad (1)$$

$$\vec{p_1} + \vec{p_i} = \vec{p_2} + \vec{p_r} \quad (2)$$

where

$$\Delta E = (E_2 - E_1)\sqrt{n^2 - 1}/n, n > 1 \quad (3)$$

is the energy change of medium (corresponding to the change of the value for rest mass of photon in medium with non-unity refractive index n), when a photon with energy $E_1$ is absorbed by medium and the photon with energy $E_2$ arises in this medium due to the interaction of absorbed photon with electron of medium that has initial energy $E_i$ and momentum $\vec{p_i}$. In (1), (2), $E_r, \vec{p_r}$ are energy and momentum of recoil electron in the corresponding Compton effect in medium with refractive index n>1 (for n<1, in all formulas n must be replaced by 1/n).

In (1), (2), we use Abraham form for momentum presentation, when

$$p_1 = E_1/cn, p_2 = E_2/cn \quad (4)$$

if n>1. Also, for electron's momentum we have

$$p_i^2 c^2 = E_i^2 - m^2 c^4, p_r^2 c^2 = E_r^2 - m^2 c^4 \quad (5)$$

If we consider the formula

$$p_r^2 = (\vec{p_i} + \vec{p_1} - \vec{p_2})^2 \quad (6)$$

on the basis of (1), (2) , it is easy to obtain the following generalization for the Compton theory:

$$(N^2-1)(E_2-E_1)^2/2mc^2n^2 - N(E_2-E_1)/n\sqrt{1-V_i^2/c^2} + (E_2\cos\theta_2 - E_1\cos\theta_1)V_i/nc\sqrt{1-V_i^2/c^2}$$
$$= E_1 E_2 (1-\cos\theta)/mc^2 n^2 \quad (7)$$

In (7), $V_i$ -is the value of initial electron velocity so that

$$p_i = mV_i/\sqrt{1-V_i^2/c^2} \quad (8)$$

$\theta, \theta_1, \theta_2$ -are the angles between $\vec{p_1}$ and $\vec{p_2}$, $\vec{p_i}$ and $\vec{p_1}$, and $\vec{p_i}$ and $\vec{p_2}$ respectively.

In (7), $N = n - \sqrt{n^2-1}$, where N<1 for all n ( n>1 and for n<1). We must also replace N on n if the value $\Delta E = 0$ in (1). Formula (7) may be presented in the form

$$(1-N^2)x^2 - 2x\left(\frac{n(N-V_i\cos\theta_2/c)}{\varepsilon\sqrt{1-V_i^2/c^2}} + 1 - \cos\theta\right) + 2(1-\cos\theta) =$$
$$= 2(\cos\theta_2 - \cos\theta_1)V_i n/c\varepsilon\sqrt{1-V_i^2/c^2} \quad (9)$$

where $x = (E_1-E_2)/E_1, \varepsilon = E_1/mc^2$. When in (9) $V_i = 0$ this formula is the same as (13) in [10] for the Compton effect in dielectric.

In the limit $(E_2-E_1)/mc^2 \ll 1$ from (9) we have:

$$-x = (E_2-E_1)/E_1 = \frac{V_i(\cos\theta_2-\cos\theta_1)n/c - \varepsilon(1-\cos\theta)\sqrt{1-V_i^2/c^2}}{(N-V_i\cos\theta_2/c)n + \varepsilon(1-\cos\theta)\sqrt{1-V_i^2/c^2}} \quad (10)$$

Thus, anomalous Compton effect is realizable if the right hand side of (10) is positive, i.e. when the followinginequalities are true:

$$n(V_i\cos\theta_2/c - N) < \varepsilon(1-\cos\theta)\sqrt{1-V_i^2/c^2} < V_i n(\cos\theta_2-\cos\theta_1)/c \quad (11)$$

or

$$n(V_i\cos\theta_2/c - N) > \varepsilon(1-\cos\theta)\sqrt{1-V_i^2/c^2} > nV_i(\cos\theta_2-\cos\theta_1)/c \quad (12)$$

From the condition providing positiveness of the left hand side of inequality (12) and its holding in principle, the restriction on the value of angle $\theta_2$ follows that coincides with the VCR realization condition obtained in [8-10] and having the form

$$1 > \cos\theta_2 > Nc/V_i ; 1 > \beta = V_i/c > N \quad (13)$$

Thus, as noted in [10], analogy of VCR effect and Compton effect is exhibited also in the form of necessary condition of realization of anomalous Compton effect in the form of inequality

(13).

Because all the momentum vectors in (2) are in the same plane (only in the special case of the medium movement this condition may not be valid), the angles $\theta, \theta_1, \theta_2$ are not independent and the value of angle $\theta_2 = \theta + \theta_1$ for some examples of electron and photon interaction (when $0 < \theta_1 < \pi/2, \theta > 0$ and $0 < \theta_2 < \pi/2$). Obviously in this case due to the positiveness of the parameter $\varepsilon > 0$, inequality (11) is not satisfied due to the negativeness of its right hand side, and the right hand side of (12), vice versa, always holds for $\cos\theta_2 < \cos\theta_1$. From the left hand side of (12),

$$\cos\theta > 1 - \frac{n(\beta\cos\theta_2 - N)}{\varepsilon\sqrt{1-\beta^2}} \tag{14}$$

where the right hand side exceeds -1 and really restricts range of $\theta$ only when holding the following complementing (13) inequality

$$\frac{2\varepsilon\sqrt{1-\beta^2}/n + N}{\beta} > \cos\theta_2 \tag{15}$$

In its turn, inequality (15) can complement inequality (13) on the value of angle $\theta_2$, when its left hand side is less than 1, i.e. under the following condition

$$\beta > \beta_1 = \frac{N\left[1 + \frac{2\varepsilon}{Nn}(4\varepsilon^2/n^2 + 1 - N^2)^{1/2}\right]}{1 + 4\varepsilon^2/n^2} > N \tag{16}$$

Inequality (16) is also take place when right hand side of (14) is smaller than -1 and we have:

$$1 > \cos\theta_2 > (N + 2\varepsilon\sqrt{1-\beta^2}/n)/\beta \tag{17}$$

when (14) give only condition $1 > \cos\theta > -1$. Thus conditions (16), (17) for anomalous Compton effect realizing have the direct analogy with VCR threshold conditions [8-10].

## 3. CONCLUSION

Thus we have considered the new theory of the Compton effect, which is based on Abraham theory for EMF in medium that needs to be experimentally verified.